\documentclass[a4paper,fleqn]{cas-sc}

\usepackage{mathtools}
\usepackage{tabu}

\usepackage[numbers,sort&compress]{natbib}

\def\tsc#1{\csdef{#1}{\textsc{\lowercase{#1}}\xspace}}
\tsc{WGM}
\tsc{QE}
\tsc{EP}
\tsc{PMS}
\tsc{BEC}
\tsc{DE}

\begin{document}
\let\WriteBookmarks\relax
\def\floatpagepagefraction{1}
\def\textpagefraction{.001}

\shorttitle{UKF for Multi-Rate Data}

\shortauthors{Ghosh et al.}

\title [mode = title]{A direct approach for full-field state-parameter estimation from fusion of noncollocated multi-rate sensor data using UKF-based algorithms}

\author[1]{Dhiraj Ghosh}[
                        orcid=0009-0003-8702-4610
                        ]


\ead{dghosh@iitk.ac.in}


\credit{Conceptualization, Methodology, Software, Writing - Original draft preparation}

\affiliation[1]{organization={Indian Institute of Technology Kanpur},
    addressline={Kalyanpur}, 
    city={Kanpur},
    postcode={UP-208016}, 
    country={India}}

\author[1]{ Adrita Kundu}


\ead{adkundu@iitk.ac.in}


\credit{Conceptualization, Methodology, Software, Writing - Original draft preparation}
    
\author[1]{ Suparno Mukhopadhyay}[
                        orcid=0000-0003-2693-762X]
                        
\cormark[1]

\cortext[cor1]{Corresponding author.}

\ead{suparno@iitk.ac.in}  
                        
\credit{Conceptualization, Methodology, Writing - Final draft preparation, Supervision, Funding acquisition and Resources}



\begin{abstract}
Heterogeneous sensor setups may entail measurements recorded at varying sampling frequencies, commonly known as multi-rate data. For system identification and state estimation with such data, existing studies mostly focus on data fusion algorithms which utilize acceleration measurements, with collocated measurements of other types obtained with lower sampling frequencies, to estimate the displacement at the collocated sensing location with the same sampling frequency as that of the acceleration measurements. The obtained displacement estimates, along with the available acceleration measurements, can then be utilized for system identification. This paper introduces a direct and straightforward methodology aimed at estimating the states (i.e., displacements and velocities) along with the unknown structural parameters from fused multi-rate data through Unscented Kalman Filter (UKF) based algorithms with a modification in the measurement update part. By utilizing all the available measurements at any time instant, which can differ due to the multi-rate nature, and by modifying the non-linear measurement equation of the dynamic system accordingly at the considered time instant, the UKF framework is suitably tailored for direct applications with multi-rate measurements. The approach is demonstrated with a variety of numerical and laboratory-scale experiments, including fusion of higher sampling frequency acceleration data with lower sampling frequency displacement, axial strain or bending strain data. The results show that the approach is successful in accurately estimating full-field states and parameters from multi-rate data. The state estimates compare well with those obtained using existing data fusion algorithms. The advantages of the approach lie in not requiring collocated sensing, in its generalizability for different types of measurements, in its simplicity and ease of implementation, and in achieving both the state and parameter estimates simultaneously.
\end{abstract}


\begin{highlights}
\item Full-field states are estimated from heterogeneous multi-rate sensor data
\item Proposed approach estimates model parameters simultaneously with states
\item Proposed fusion technique does not need collocated sensors 
\item Same or different types of multi-rate data can be used within the same framework 
\end{highlights}


\maketitle

\section{Introduction}

For vibration-based structural health monitoring purposes, several advancements have been made in sensor technologies, leading to a wide variety of sensors for capturing different types of vibration measurements. Acceleration measurements are the most commonly utilized as they contain information about the overall global dynamic behaviour of structures. However, they do not capture lower frequency contents well. Displacement measurements, on the other hand, can better capture the lower frequency contents~\cite{smyth2007multi,feng2016vision,jana2021computer}.
Additionally, strain measurements offer insights into local deformations and are very useful for damage detection scenarios~\cite{li2010hypersensitivity}. Therefore,  the fusion of acceleration with displacement/strain measurements can be beneficial.
However, the sampling frequencies of acceleration measurements are often high compared to displacement/strain measurements. Additionally, not only may different types of measurements have varying sampling frequencies, but similar types of measurements can also differ in sampling frequency based on the sensor type (e.g. LVDT versus camera, strain gauges versus strain obtained using digital image correlation, etc).
Thus, in state-parameter estimation using same/different types of data, measured at different sampling frequencies, appropriate fusion of the measured data is necessary, addressing the challenges posed by the differences in sampling frequencies.


In the existing literature, this topic is typically sub-divided into two separate albeit connected research objectives: (a) Fusion of multi-rate measurements of acceleration with either displacement~\cite{smyth2007multi,kim2014autonomous,sohn2016dynamic,zheng2019data,pal2024data,xu2017long,ma2022real,he2023baseline} or strain~\cite{sarwar2020bridge,he2022displacement} at a collocated location, in order to obtain the corresponding displacement estimate at the higher sampling frequency of the acceleration measurement; and (b) Structural system identification and states (displacements and velocities) estimation, with acceleration and displacement measurements while assuming that all measurements (not necessarily collocated) have the same sampling frequencies~\cite{chatzi2009unscented,chatzis2015experimental}. In the first objective, the acceleration-displacement fusion methods solely exploit the derivative-based relationships between displacement, velocity and acceleration at the collocated location, with many of these methods employing some version of the Kalman filtering framework~\cite{smyth2007multi,kim2014autonomous,sohn2016dynamic,zheng2019data,pal2024data}. For the acceleration-strain fusion methods, strain-based displacements are first obtained with/without using the system properties of the structure such as mode shapes or finite element model, which is followed by the utilization of the aforementioned derivative-based relationships to estimate the displacement at the collocated location. Once the displacements at the collocated sensor locations are obtained in the higher sampling frequency of the accelerations, these data (accelerations and displacements) can then be used for structural parameter estimation and/or damage detection as a second step. It would be more advantageous if: (a) the estimation of the states (full-field displacements and velocities) and structural system identification can be done simultaneously using different types of multi-rate measurements, (b) the requirement of collocated sensing can be relaxed, and (c) the approach is generalized in the sense that it can handle multi-rate data of same or different types within the same framework (instead of having different frameworks for strains, displacements etc.). The above requirements define the scope and contribution of this paper.



To address the challenges of state estimation and system identification with multi-rate data, the problem statement can be cast as a Bayesian state-parameter estimation problem, wherein the multi-rate measurements may be directly utilized for providing the estimates of the states and the structural parameters simultaneously. To this end, various Bayesian filtration techniques such as extended Kalman filters, unscented Kalman filters (UKF), particle filters etc. can be used. As the states' estimates will also be dependent on the physics of the structure, collocated sensing locations may not be required, 
which is a necessity for the existing data fusion methods providing displacement estimates at the collocated sensing locations only. Additionally, because of the physical relationship between the structure and its dynamic states, the estimation of states at all locations of the structure may also be possible with measurements from limited instrumentations. In this work, such a Bayesian state-parameter estimation algorithm tailored for multi-rate measurements has been proposed based on the UKF algorithm, a popular technique which has been utilized for state-parameter estimation of a wide variety of structural systems~\cite{chatzi2009unscented,chatzis2015experimental,mariani2007unscented,wu2008real,chatzi2010experimental,xie2012real,erazo2019bayesian,lei2019novel}. The proposed approach directly utilizes the measurements without any additional pre-processing with data fusion algorithms, and also does not require collocated sensor pairs. The algorithm can utilize various types of multi-rate measurements, and provide estimates of the states at all degrees of freedom (DOFs), i.e., full-field estimation of displacement and velocity, along with the estimates of the structural parameters as well. Suitable numerical illustrations and experimental applications have been shown to demonstrate the efficiency of the proposed methodology with multi-rate acceleration-displacement and multi-rate acceleration-strain measurements.


\section{Methodology}
\begin{figure}[b!]
	\centering
		\includegraphics[scale=.65]{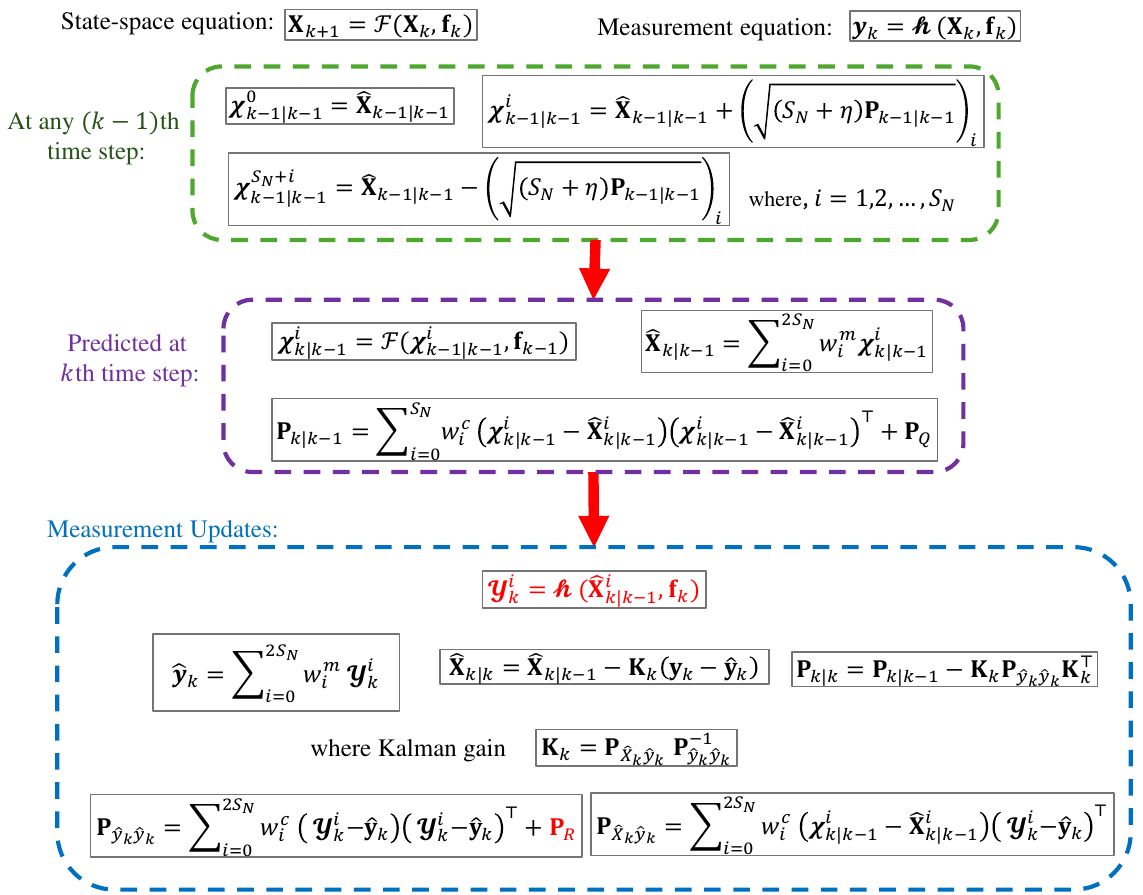}
	   \caption{Flowchart of the UKF algorithm} 
	\label{FIG:UKF_algo}
\end{figure}

{\label{sec:UKF framework}}
The joint state-parameter, considering displacement and velocity as the states, can be estimated by using the UKF algorithm~\cite{julier2004unscented,chatzi2009unscented}. A flowchart of the UKF algorithm has been shown for a non-linear state-space equation for a general dynamic system in Fig.~\ref{FIG:UKF_algo}.
Here, $\mathbf{X}$ is the state vector which is augmented with the structural parameters, and $\hat{\mathbf{X}}$ is the estimated augmented state vector.
The input is represented as $\mathbf{f}$, and measurements are denoted as $\mathbf{y}$. The size of the augmented state vector is indicated by $S_N$, and the total number of sigma points $\boldsymbol{\chi}^i$ becomes $2S_N+1$. The spread of the sigma points is controlled by the scaling parameter $\eta$, and in the flowchart, $\left(\sqrt{(S_N+1)\mathbf{P_{k-1|k-1}}}\right)_i$ indicates the $i$th column of the matrix $\sqrt{(S_N+1)\mathbf{P_{k-1|k-1}}}$. $\mathbf{P}_Q$ and $\mathbf{P}_R$ are the covariance matrices for zero-mean process and measurement noise, respectively. The weights associated with the sigma points can be different for computing mean and covariance, which are denoted as w$_i^m$ and w$_i^c$, respectively. In the flowchart shown in Fig.~\ref{FIG:UKF_algo}, $(.)_{k|k-1}$ and $(.)_{k|k}$ are the a priori and a posteriori estimates, respectively, at the $k$th time step.

The proposed algorithm directly incorporates multi-rate data in the "measurement updates" step in the flowchart of UKF (as shown in Fig.~\ref{FIG:UKF_algo}), while using the measurements to update the state predictions.
Based on available measurements at that specific time step, the "measurement updates" step is the only part that varies. In this step, $\boldsymbol{\mathscr{Y}}$ obtained using the~measurement~function $\boldsymbol{\mathscr{h}}$ will contain only the rows corresponding to the sensor type active during each time step. The covariance matrix of measurement noise is also changed accordingly, based on the sensor types present at that time step. 

For illustration, consider an N-degree of freedom (DOF) system which is instrumented with a total of $N_S$ number of sensors of different types. Vibration data recorded using the $i$th sensor, i.e., $\widetilde{S}_i$ has the sampling frequency of $\mathcal{F}_i$ where $i=1,2,\cdots,N_S$. At the $k$th time step, if measurements from a subset of sensors $\mathcal{S_K}$ containing $\mathcal{J}_k$ number of sensors ($\mathcal{J}_k\leq N_S$) are available, the measurement function $\boldsymbol{\mathscr{h}}$ will give the responses corresponding to only those sensors in the subset $\mathcal{S_K}$ for that particular time step. Consequently, the measurement noise covariance matrix will become of size $\mathcal{J}_k\times\mathcal{J}_k$ according to the sensors contained in the subset $\mathcal{S_K}$. For instance, if sensors ${\widetilde{S}_1}$, ${\widetilde{S}_4}$, and ${\widetilde{S}_7}$ are available at $k$th time step, i.e., $\mathcal{S_K}=\{\widetilde{S}_1,\widetilde{S}_4,\widetilde{S}_7\}$ where $\mathcal{J}_k$ is equal to 3, then the measurement function $\boldsymbol{\mathscr{h}}$ will be:
\begin{equation}\label{EQN: measurement}
    \boldsymbol{\mathscr{h}}= \left[ \mathscr{h}_{\widetilde{S}_1}, \mathscr{h}_{\widetilde{S}_4}, \mathscr{h}_{\widetilde{S}_7} \right]^\top
\end{equation}
The measurement noise covariance matrix for that time step will be updated as: 
\begin{equation}\label{EQN: measurement noise matrix}
    \mathbf{P}_R = \textrm{ diag}\left(\left[ \sigma^2_{\widetilde{S}_1}, \sigma^2_{\widetilde{S}_4}, \sigma^2_{\widetilde{S}_7} \right] \right)
\end{equation}
where $\text{diag}\left(\mathbf{A}\right)$ for vector $\mathbf{A}$ creates a diagonal matrix with all the elements in $\mathbf{A}$, and
$\sigma_{\widetilde{S}_i}^2$ is the variance of the measurement noise for the $i$th sensor.

\par
This above approach can be used with any type of UKF-based algorithm. Hence, for illustration purposes, in addition to the standard UKF, the CGUKF~\cite{li2022constrained} is also considered in this paper. The CGUKF belongs to the class of constrained UKF algorithms, where apriori physics-based information (e.g., non-negative stiffness) about the structure is imposed in the form of constraints, in order to improve the performance of the UKF algorithm. In the CGUKF algorithm, these constraints are incorporated during the Kalman gain $\mathbf{K}_k$ computation. The gain $\mathbf{K}_k$ is computed as the solution to a constrained optimization problem, minimizing the posterior state covariance matrix subject to available physics-based constraints on the states and/or parameters.
For adapting the CGUKF algorithm, the proposed algorithm incorporates multi-rate data in the "measurement updates" step.
The function $\boldsymbol{\mathscr{h}}$ and the measurement noise covariance matrix $\mathbf{P}_R$ will be changed for multi-rate data in the same manner as shown in Eqs.~\ref{EQN: measurement}~and~\ref{EQN: measurement noise matrix}. Other UKF-based algorithms can also be tailored in a similar way for multi-rate data.

\section{Numerical Illustration} \label{sec:num}
To illustrate the proposed methodology, two different types of structures have been considered. First, a two story shear building is considered where the acceleration measurement is fused with displacement measurement recorded at a lower sampling frequency than the acceleration. Then, the fusion of acceleration is done with strain measurements of lower sampling frequency for a 2D truss model.
In all the examples, about 50\% of the active DOFs/members are considered to be instrumented, so to avoid issues related to observability/identifiability of the states/parameters. While even lesser instrumentation may be used in some cases, that would require considering observability/identifiability of states and parameters and optimally locating of such sensors, which require exclusive treatment~\cite{udwadia1994methodology, papadimitriou2004optimal, ghosh2022fisher, ghosh2024fisher, mukhopadhyay2015structural, mukhopadhyay2016structural, chatzis2015observability}. 

\subsection{Two-Storied Shear Building} \label{sec:2-storied frame}

\begin{figure}
	\centering
		\includegraphics[scale=.61]{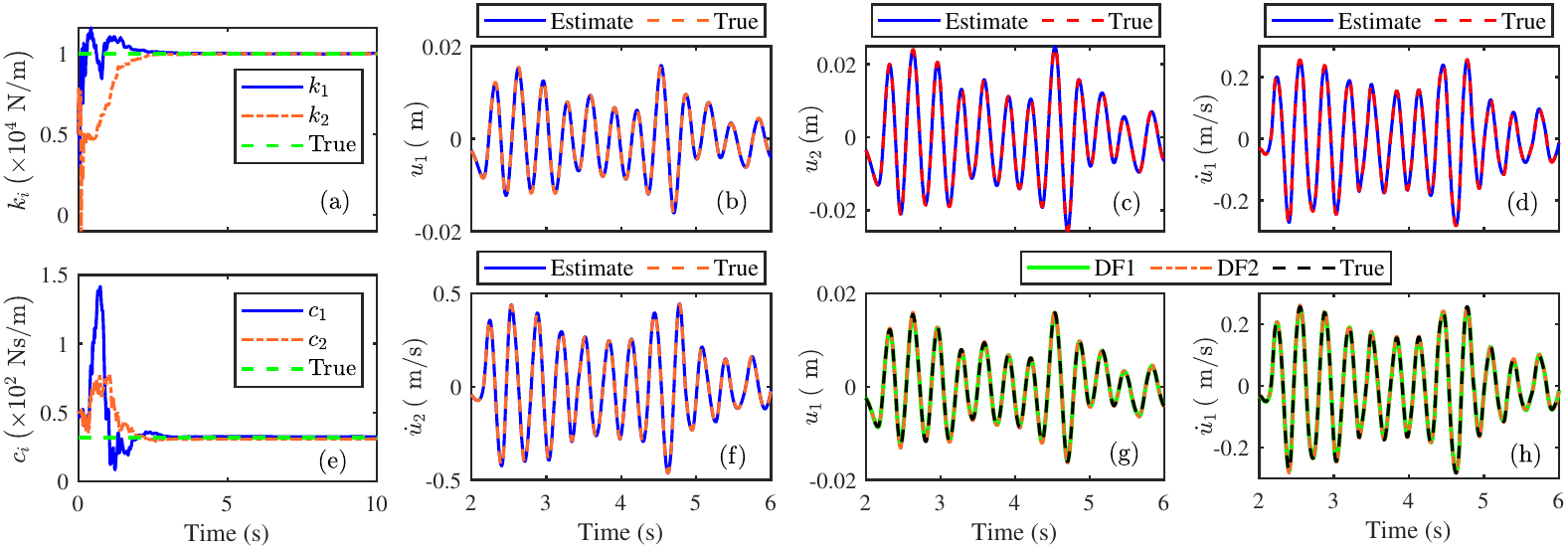}
	   \caption{(a-f) State-parameter estimates obtained by the proposed approach, and (g-h) Performance of conventional data fusions algorithms, for the two-storey shear frame numerical example}
	\label{FIG:2 storied frame}
\end{figure}

\begin{figure}
	\centering
		\includegraphics[scale=.61]{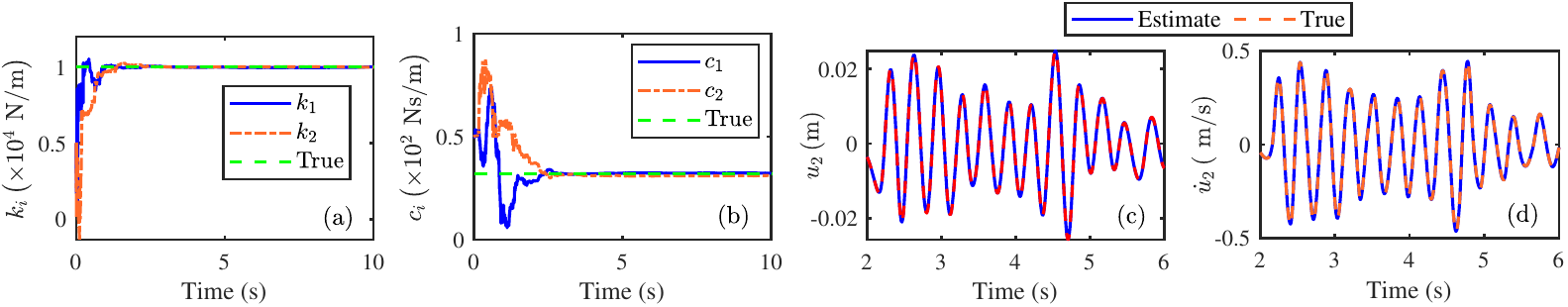}
	   \caption{State-parameter estimates obtained with the sampling frequencies of acceleration and displacement measurements not being each others multiple, for the two-storey shear frame numerical example}
	\label{FIG:2 storied frame 2}
\end{figure}

The structure is modelled as a two DOF shear frame with the following structural parameters: floor masses $m_1=m_2=10 \text{ kg}$, story stiffnesses $k_1=k_2=10 \text{ kN/m}$, and damping coefficients $c_1=c_2=32 \text{ Ns/m}$, where the subscripts 1 and 2 denote the first and second floors, respectively. The frame is subjected to the El Centro ground motion. It is considered that the measurements from the structure consist of noisy acceleration and displacement from the first floor only, which are sampled at 500 Hz and 50 Hz, respectively, and corrupted with additive Gaussian white noise (GWN) of 10\% root mean squared (RMS) noise to signal ratio. For this example, owing to the sampling frequencies, at any time instant, the measured data can consist of: (a) only acceleration measurement, and (b) both displacement and acceleration measurements. The measurement function $\boldsymbol{\mathscr{h}}$ and the measurement noise covariance matrix $\mathbf{P}_R$ are modified accordingly during every instance.
 For example, at time $t=0.398$ s (only acceleration measurement available), the measurement equation and its corresponding measurement noise covariance matrix will become: $\mathbf{y}_k = \ddot{u}_1 $ and $\mathbf{P}_R = \sigma^2_{\ddot{u}_1}$, respectively. At time $t=0.4$ s when both acceleration and displacement measurements are available, these equations change, as follows: $\mathbf{y}_k = \big[ \ddot{u}_1, {u}_1\big]^\top$ and $\mathbf{P}_R = \textrm{diag}\left( \big[ \sigma^2_{\ddot{u}_1}, \sigma^2_{{u}_1}\big] \right)$, respectively. In these equations, $\ddot{u}_i$ and $u_j$ are the acceleration of the $i$th floor and displacement of the $j$th floor, respectively.

Assuming the floor masses to be known, the UKF algorithm with the proposed modification in the measurement update is employed to perform estimation of the state-parameter vector $\mathbf{X} = \left\{u_1,u_2,\dot{u}_1,\dot{u}_2,k_1,k_2,c_1,c_2\right\}^\textrm{T}$, where $\dot{u}_i$ is the floor velocity of the $i$th floor $,\forall i=1,2$. The unknown structural parameters converge quite accurately to their true values, as depicted in Figs.~\ref{FIG:2 storied frame}a and e. The estimation of the states: $u_i$ and $\dot{u}_i,\forall i=1,2$, are shown in Figs.~\ref{FIG:2 storied frame}b-d and~\ref{FIG:2 storied frame}f. It is observed that the states are estimated with high accuracies, with RMS errors (compared to true values) below 2\%. 

\begin{figure}[b!]
	\centering
		\includegraphics[scale=.37]{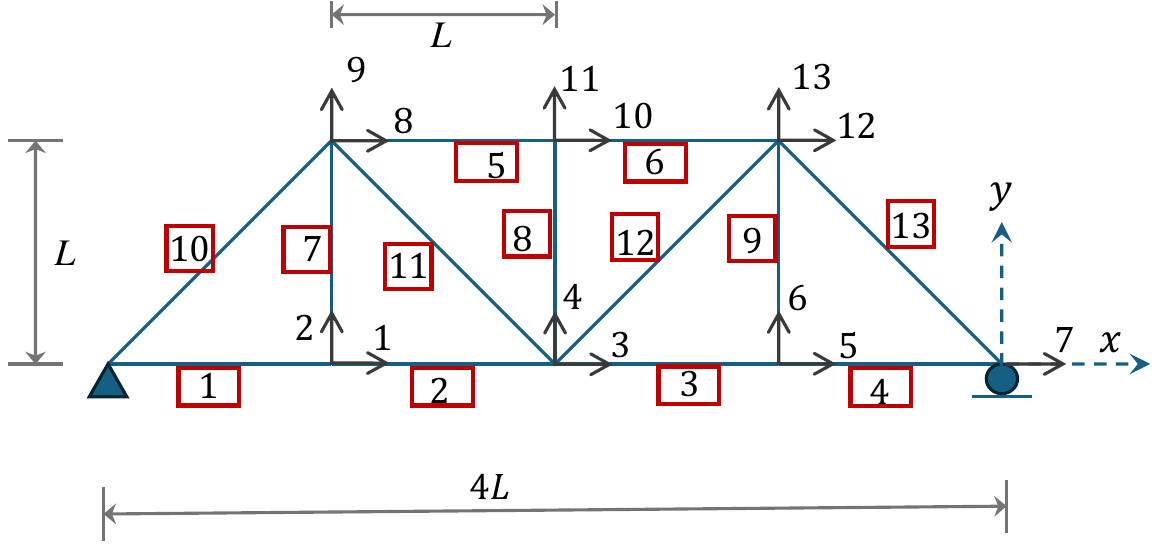}
	   \caption{Schematic diagram of the 2D truss with DOFs and element numbers (inside boxes) }
	\label{FIG:2Dtruss}
\end{figure}
For comparison, the estimation of displacement $u_1$ and velocity $\dot{u}_1$ at the collocated sensor location i.e., floor 1, is also performed with two conventional multi-rate data fusion algorithms (which require collocation): (a) Rauch-Tung-Striebel algorithm-based smoothing Kalman filter~\cite{smyth2007multi}, and (b) Short-term memory Kalman filter~\cite{pal2024data}. For brevity, these algorithms are referred to as DF1 and DF2, respectively, and the estimates provided by them are displayed in Figs.~\ref{FIG:2 storied frame}g-h. Their performances are quite good as well, with RMS errors (compared to true values) of $u_1$ and $\dot{u}_1$ between 4\% and 5\%. Comparing the performance of the proposed approach with DF1 and DF2 algorithms, it is evident that the proposed approach performs as good as the existing methods, while having the advantages of simultaneous parameter estimation, non-requirement of collocated sensing, and estimation of unmeasured states.


The estimation is also performed with the acceleration and displacement measurements having sampling frequencies of 500 Hz and 30 Hz, respectively. The estimation results are good and very similar to that seen previously, as observed from the estimates of the parameters and the states $u_2$ and $\dot{u}_2$ shown in Fig.~\ref{FIG:2 storied frame 2}. This example denotes a case where the sampling frequencies of different types of measurements are not multiples of each other. This introduces an additional type of measurement scenario where only the displacement measurement is present, which is also handled by modifying $\boldsymbol{\mathscr{h}}$ and $\mathbf{P}_R$ suitably. Furthermore, in such cases, the size of the time step is not constant and not always equal to that of the higher frequency acceleration measurements, as the displacement measurement may fall between two acceleration measurements. For example, let two consecutive acceleration measurements be at times $t_{a1}=1.032$ s and $t_{a2}=1.034$ s, and the middle displacement measurement be at time $t_{d} = 1.0\overline{3}$ s. After the state-parameter update with only acceleration measurements at time $t_{a1}$, the next update is with only displacement measurements at time $t_{d}$, having a time step of $t_d - t_{a1} = 0.001\overline{3}$ s, followed by the subsequent update with only acceleration measurements at time $t_{a2}$, while having a time step of $t_{a2} - t_{d} = 0.000\overline{6}$ s.



\subsection{2D Truss} \label{sec:truss}
In this example, a 2D truss as shown in Fig.~\ref{FIG:2Dtruss} is considered.
The unrestrained degrees of freedom (DOFs) and member numbers of the truss are as marked in Fig.~\ref{FIG:2Dtruss}. The cross-sectional areas of the top, bottom, diagonal and vertical members are 80, 100, 90 and 60 mm$^2$, respectively. The modulus of elasticity is 200 GPa, and the material mass density is 7850 kg/m³. Additionally, a lumped mass of 10 kg is added at all the unrestrained DOFs except for DOF 7, where a 5 kg mass is added. Rayleigh damping is adopted considering 2\% modal damping ratio for the first nine modes. A GWN is applied as input at DOF 2. Acceleration responses are generated at a sampling frequency of 1000 Hz, and axial strains of the members are generated at a sampling frequency of 250 Hz. Additionally, 5\% and 10\% RMS GWN are added to the generated acceleration and strain responses, respectively, to get the corresponding measured responses. To illustrate the proposed methodology, limited number of sensors are considered: acceleration measurements at horizontal DOFs: 7-8-12, and vertical DOFs: 2-9-11, and strain measurements on members:~3-7-8-9-10-12-13.

\begin{figure}[t!]
	\centering
		\includegraphics[scale=.625]{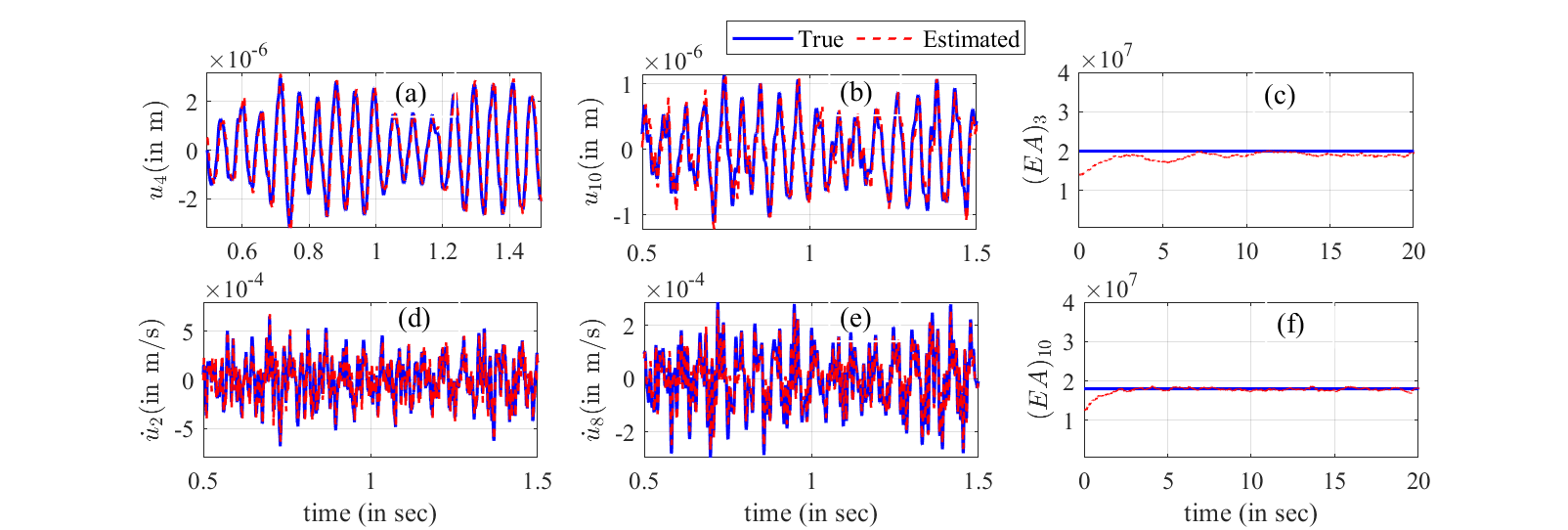}
	   \caption{Comparison of true and estimated displacement at (a) DOF 4 (RMS error 9.73\%) and (b) DOF 10 (RMS error 0.93\%), and velocity at (d) DOF 2 (RMS error 2.01\%) and (e) DOF 8 (RMS error 1.85\%), along with the convergence of estimated axial rigidity of (c) member 3 and (f) member 10, for the 2D-truss numerical example}
	\label{FIG:mR_est}
\end{figure}

In this example, CGUKF is used to estimate the states and parameters. The augmented state-parameter vector considered is $\mathbf{X} = \{u_1,\cdots,u_{13},\dot{u}_1,\cdots,\dot{u}_{13},EA_1,\cdots,EA_{13},\alpha,\beta \}^\textrm{T}$, where $(EA)_i$ is the axial rigidity of $i$th member of the 2D truss, and $\alpha$ and $\beta$ are Rayleigh's damping coefficients.
In the "measurement updates" step, the measurement equation and its corresponding measurement noise covariance matrix will change, as follows:
\begin{equation}\label{eq:yPr_truss}
 \begin{matrix}
 \begin{matrix}
\textrm{ acceleration-strain} \\ 
\textrm{ (fused) measurement}     
 \end{matrix}
 :
 \begin{cases}
    \mathbf{y}_k = \bigg[
        \ddot{u}_2, \ddot{u}_7, \ddot{u}_8, \ddot{u}_9, \ddot{u}_{11}, \ddot{u}_{12}, \epsilon_3, \epsilon_7, \epsilon_8,\epsilon_9, \epsilon_{10}, \epsilon_{12}, \epsilon_{13}\bigg]^\top 
        \\
    \mathbf{P}_R = \textrm{diag}\left( \bigg[ \sigma^2_{\ddot{u}_2}, \sigma^2_{\ddot{u}_7}, \sigma^2_{\ddot{u}_8},\sigma^2_{\ddot{u}_9}, \sigma^2_{\ddot{u}_{11}}, \sigma^2_{\ddot{u}_{12}}, \sigma^2_{\epsilon_3}, \sigma^2_{\epsilon_7} , \sigma^2_{\epsilon_8}, \sigma^2_{\epsilon_9}, \sigma^2_{\epsilon_{10}} , \sigma^2_{\epsilon_{12}}, \sigma^2_{\epsilon_{13}}  \bigg] \right)
    \end{cases} 
    \\
 \begin{matrix}
\textrm{ acceleration-only} \\ 
\textrm{ measurement}     
 \end{matrix}
 :
     \begin{cases}
    \mathbf{y}_k = \bigg[
        \sigma^2_{\ddot{u}_2}, \sigma^2_{\ddot{u}_7}, \sigma^2_{\ddot{u}_8},\sigma^2_{\ddot{u}_9}, \sigma^2_{\ddot{u}_{11}}, \sigma^2_{\ddot{u}_{12}} \bigg]^\top \\
    \mathbf{P}_R = \textrm{diag}\left( \bigg[ \sigma^2_{\ddot{u}_2}, \sigma^2_{\ddot{u}_7}, \sigma^2_{\ddot{u}_8},\sigma^2_{\ddot{u}_9}, \sigma^2_{\ddot{u}_{11}}, \sigma^2_{\ddot{u}_{12}} \bigg] \right)
    \end{cases}
 \end{matrix}
\end{equation}
where $\epsilon_j$ is the strain measurement of the $j$th member of the truss.

\begin{figure}[t!]
	\centering
		\includegraphics[scale=.75]{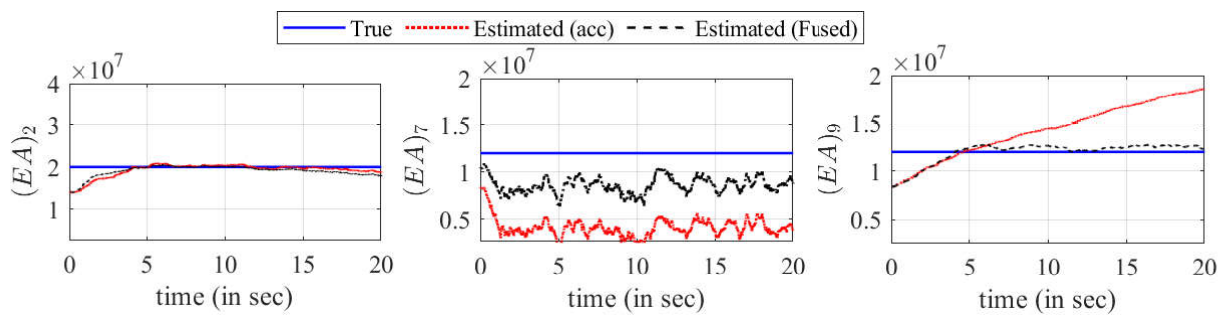}
    \caption{Comparison of the axial rigidity estimates using only acceleration measurements with those obtained from fused data, for members: (a) 2, (b) 7, and (c) 9, for the 2D-truss numerical example}
	\label{FIG:mR_est_acc}
\end{figure}

\begin{figure}[t!]
	\centering
		\includegraphics[scale=.70]{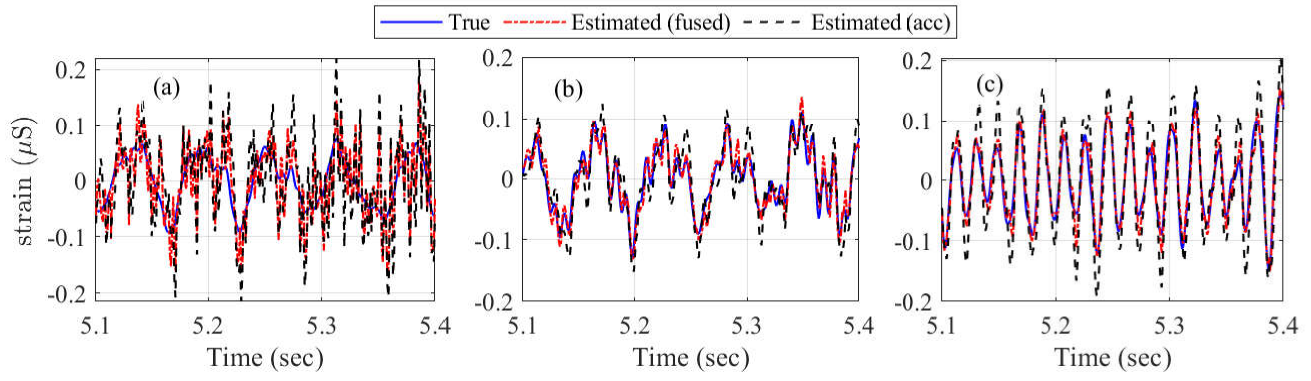}
    \caption{Comparison of the axial strains estimates using only acceleration measurements with those obtained from fused data along with the true values, for members: (a) 2, (b) 5, and (c) 9, of the truss example}
	\label{FIG:mR_est_strain}
\end{figure}

It is observed that the estimates of states and structural parameters (axial rigidity of the members) match their true values well. For instance, comparisons between estimated and true values for one vertical ($u_4$) and one horizontal ($u_{10}$) displacement are depicted in Fig.~\ref{FIG:mR_est}a and b, respectively. Additionally, the estimated velocities in both vertical and horizontal directions are compared with their~true~values~in~Fig.~\ref{FIG:mR_est}d~and~e,~respectively.
Also, the identified parameters agree well with their true values. For instance, the convergences of two identified parameters, namely the axial rigidity of members 3 and 10 of the 2D truss, are shown in Fig.~\ref{FIG:mR_est}c and f, respectively.
The ratio of estimated parameters to the true parameters are shown in Table~\ref{tab:2D_truss} for both fused data and acceleration only identification.
Note that, in this example, using only acceleration measurements does not suffice to accurately identify the axial rigidity of all members; the results either diverge or converge incorrectly, as shown in Fig.~\ref{FIG:mR_est_acc}. From the comparison of the identification using acceleration only with the fused data, it is seen that $(EA)_7$ is converging incorrectly (Fig.~\ref{FIG:mR_est_acc}b), and $(EA)_9$ (Fig.~\ref{FIG:mR_est_acc}c) is diverging, in the case of limited number of acceleration only data. This can be possibly attributed to observability/identifiability issues. Detailed investigation on state/parameter observability/identifiability is not addressed in this work, as they require an exclusive treatment, as can be found in~\cite{mukhopadhyay2015structural, mukhopadhyay2016structural, chatzis2015observability}, and is beyond the scope of the current work. A comparison of the estimated strains obtained using the displacement-strain transformation matrix from the identified full-field displacement is presented in Fig.~\ref{FIG:mR_est_strain}. It is observed that strain estimates derived from acceleration-only data exhibit greater discrepancies compared to the strain estimates from fused data.
To successfully determine the axial rigidity of all members, more accelerometers must be added, or different types of responses (in this case, axial strains) need to be measured. However, measuring strains, as opposed to accelerations, is typically more cost-effective, illustrating the utility of heterogeneous sensing and the proposed approach in structural identification.

\begin{table}
    \centering
\caption{Ratio of estimated axial rigidities to the true parameters in both fused data and acceleration only data scenarios}
    \label{tab:2D_truss}
    \begin{tabular}{c|c|c}
          Member number & Fused data & Acceleration only  \\ \hline
          1 &  1.26&     1.47\\
 2& 0.89&0.93\\
 3& 0.98&0.78\\
 4& 0.80&0.65\\
 5& 0.94&1.13\\
 6& 0.86&0.80\\
 7& 0.79&0.35\\
 8& 0.97&1.11\\
 9& 1.02&1.55\\
 10& 1.00&1.26\\
 11& 0.75&0.50\\
 12& 1.00&1.12\\
 13& 0.99&1.23\\ \hline
 \end{tabular}
    \end{table}

\section{Experimetal Validation}

The proposed methodology is now validated with two laboratory-scale experimental applications considering fusion of different types of lower-frequency measurements with high-frequency acceleration measurements.
In these examples also, about 50\% of the active DOFs/elements have been instrumented to avoid issues related to observability/identifiability of the states/parameters.

\subsection{Four-Storied Laboratory-Scale Steel Frame}

\begin{figure}[t!]
	\centering
		\includegraphics[scale=.425]{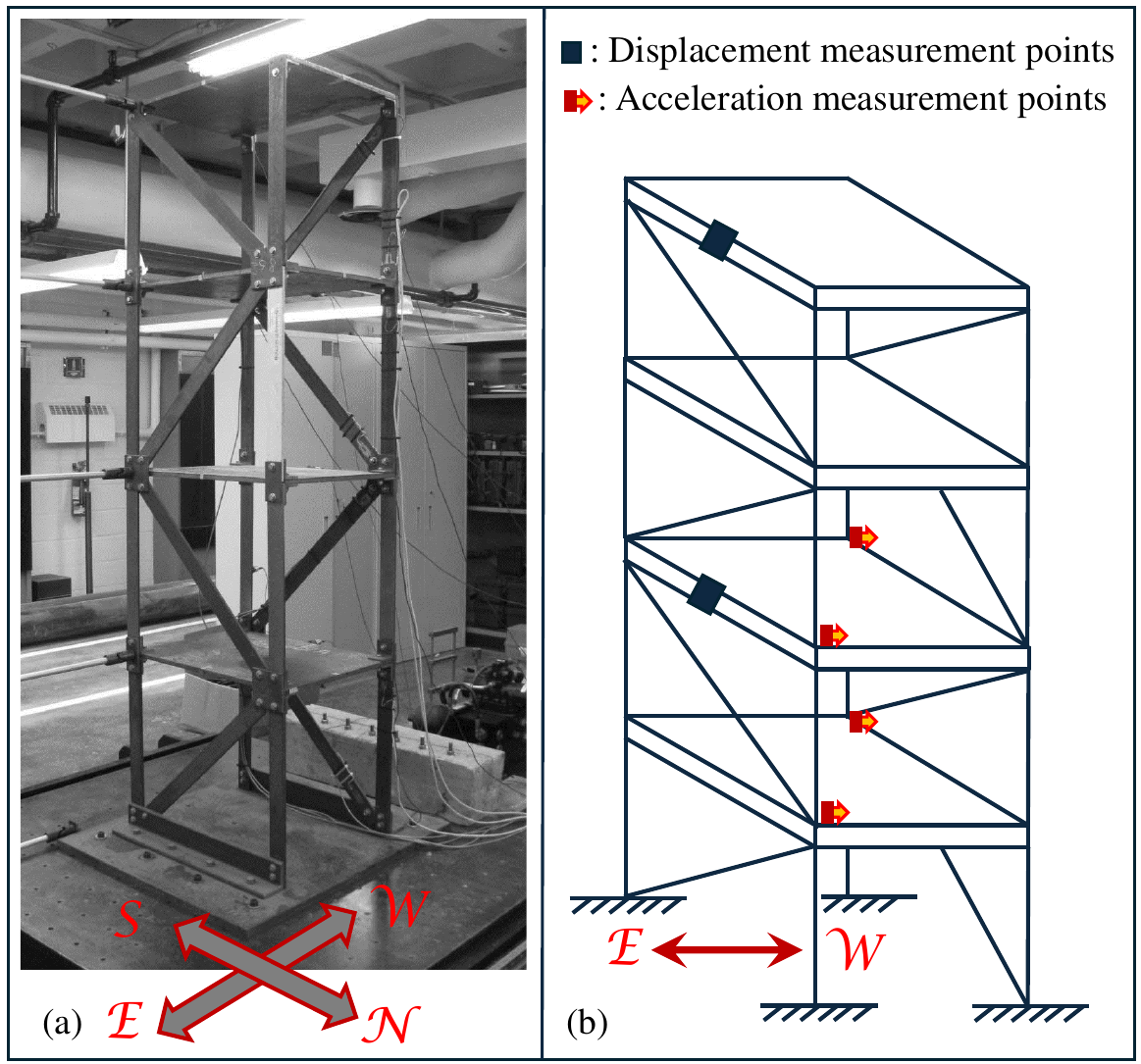}
	   \caption{(a) Four-storied steel frame, and (b) schematic diagram of the frame depicting the locations of the considered accelerometers and points of laser displacement measurements.}
	\label{FIG:smy_exp}
\end{figure}

The first application is with a four-storied shear-type steel frame, which is subjected to ground motions in the East-West direction (weaker in bending) on the hydraulic uniaxial shake table of Columbia University, New York (Fig.~\ref{FIG:smy_exp}a). Detailed information about the experimental setup can be found in~\cite{mukhopadhyay2015structural,chatzis2015experimental}. The instrumentation consists of multiple accelerometers on every floor, along with four laser sensors to measure the absolute displacement at the centre of each floor~\cite{chatzis2015experimental}. The displacement and acceleration of the shake table are measured with the actuator's inbuilt LVDT and a MEMs accelerometer, respectively. The absolute accelerations and displacements of each floor and the base are obtained with a sampling frequency of 200 Hz. The measurements are quite clean having low noise content. In the example considered in this section, the responses obtained from the frame when subjected to the EC8 ground motion are utilized. It is considered that accelerations are collected at floors 1 and 2 only, while displacements are collected at floors 2 and 4 only. The displacement measurements are further downsampled to 50 Hz. The considered accelerometers and the points of laser displacement measurements are depicted in a schematic diagram in Fig.~\ref{FIG:smy_exp}b. The absolute acceleration measurements for each of the considered floors are taken by averaging the measurements from the two accelerometers of that particular floor. The relative displacement measurements for the considered floors are taken by subtracting the shake-table LVDT displacement from the absolute displacements of the floors obtained with the laser sensors.

The measurement noise covariance matrix, $\mathbf{P}_R$, is assumed to be diagonal under the premise that the measurement noises of different sensors are uncorrelated. The values on the diagonal elements of the matrix $\mathbf{P}_R$ vary in magnitude depending on their corresponding sensors. For a particular sensor, first, the standard deviation from the data recorded before the application of the input force is computed. Then, the squared value of this standard deviation is used as the variance of the measurement noise in that sensor. The variances of all sensors computed in this manner are used as their respective diagonal elements of the $\mathbf{P}_R$ matrix.

\begin{figure}[t!]
	\centering
		\includegraphics[scale=.62]{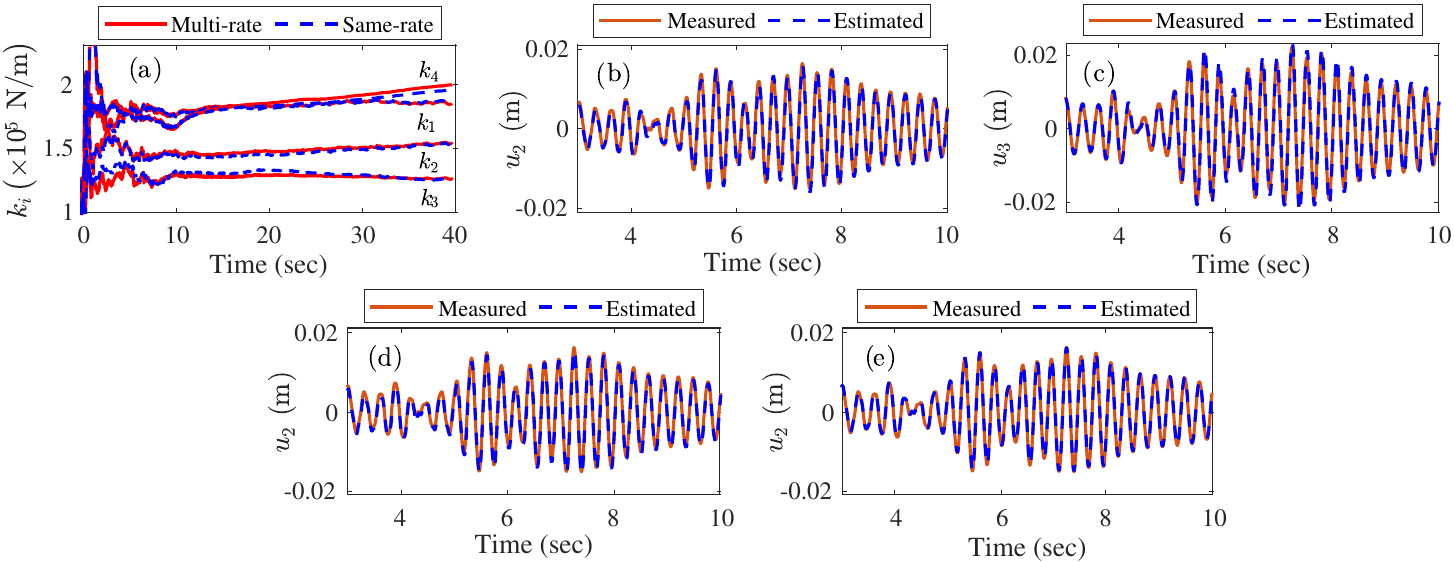}
	   \caption{Estimates for experimental steel frame: (a-c) estimates with the proposed approach: story stiffnesses $k_i$, $u_2$, and $u_3$, respectively, (d-e) estimated story displacement $u_2$ for DF1 and DF2, respectively.}
	\label{FIG:Chatzis_exp}
\end{figure}

The dynamic model of a 2D four-storied shear frame considering Rayleigh damping is adopted as the underlying numerical model. The state-parameter estimation algorithm employed is the standard UKF, with the state-parameter vector $\mathbf{X} = \left\{u_1,\cdots,u_4,\dot{u}_1,\cdots,\dot{u}_4,k_1,\cdots,k_4,\alpha,\beta\right\}^\textrm{T}$, where $u_i$ and $\dot{u}_i$ are the relative floor displacements and relative floor velocities of the $i$th floor, $\forall i = 1,\cdots,4$, $k_i$ is the stiffness of the $i$th floor, and $\alpha$ and $\beta$ are the Rayleigh's damping coefficients. The estimated stiffnesses are depicted in Fig.~\ref{FIG:Chatzis_exp}a, and are also compared with those obtained from same-rate fused data (when the displacement measurements are not downsampled and their sampling frequencies are also 200 Hz). The stiffnesses obtained with both types of data fusions lead to similar results. The RMS deviations of all the estimated relative displacements $u_i,\ \forall i=1,\ldots,4$ with respect to their actual measurements are given in row 1 of Table~\ref{TABLE: Displacement Errors}. They can be seen to be in the range of 6-8 \%. The estimated relative displacements of floors 2 and 3 are shown in Fig.~\ref{FIG:Chatzis_exp}b and c, respectively, and also compared with their respective measurements. The estimated and measured quantities are seen to agree quite well with each other. It is to be noted that the displacement of floor 3 is not utilized by the proposed approach as a measurement, yet the estimation of this displacement is quite accurate.

For comparing the state estimation performance with the conventional data fusion algorithms at the collocated sensing location (floor 2), $u_2$ is also estimated with DF1 and DF2 algorithms. However, these algorithms work with those accelerations and displacement at the collocated location only, which are measured from the same reference frame. Hence, the absolute displacement measurements from the laser sensor, assumed to be obtained with the sampling frequency of 50 Hz, is utilized. DF1 and DF2, therefore, give estimates of the absolute displacement at floor 2. The estimated relative displacement $u_2$ is then obtained by subtracting the shake-table LVDT displacement measurement from the estimated absolute displacement. These estimates of $u_2$ thus obtained with DF1 and DF2 algorithms are compared with their respective measurements and shown in Fig. 7d and e, respectively. The RMS deviations of the estimated $u_2$ with respect to the actual measurements are shown in Table~\ref{TABLE: Displacement Errors} as well. The RMS deviations are found to be very low which can be attributed to the very low noise content in the measurements. The relatively higher RMS deviations in the case of the proposed approach can be possibly owed to modelling inaccuracies.

The proposed methodology is also tested with a set of non-collocated multi-rate data: accelerations from floors 1 and 2, and displacements from floors 3 and 4. The RMS deviations of the obtained estimates of the relative floor displacements are also provided in Table~\ref{TABLE: Displacement Errors}. It can be seen that the results are very similar to those obtained before with the set of measurements having collocated sensing at floor 2. 


\begin{table}[t!]
\centering
\caption{RMS deviations from measured data (in \%) of relative displacement estimates obtained by the proposed approach, DF1 and DF2 algorithms}
\label{TABLE: Displacement Errors}
\begin{tabu}{c|cccc}
\tabucline[2pt]{-}
Displacement & $u_1$ & $u_2$ & $u_3$ & $u_4$ \\
\hline
Proposed approach (collocated at floor 2) & 7.29 & 7.33 & 7.84 & 6.5\\ 
DF1 (collocated at floor 2)& - & 2.03 & - & -\\ 
DF2 (collocated at floor 2)& - & 1.72 & - & -\\ 
Proposed approach (non-collocated)  & 7.04 & 7.26 & 6.29 & 6.5\\ 
\tabucline[1.5pt]{-}
\end{tabu}
\end{table}

\begin{figure}
	\centering
		\includegraphics[scale=.55]{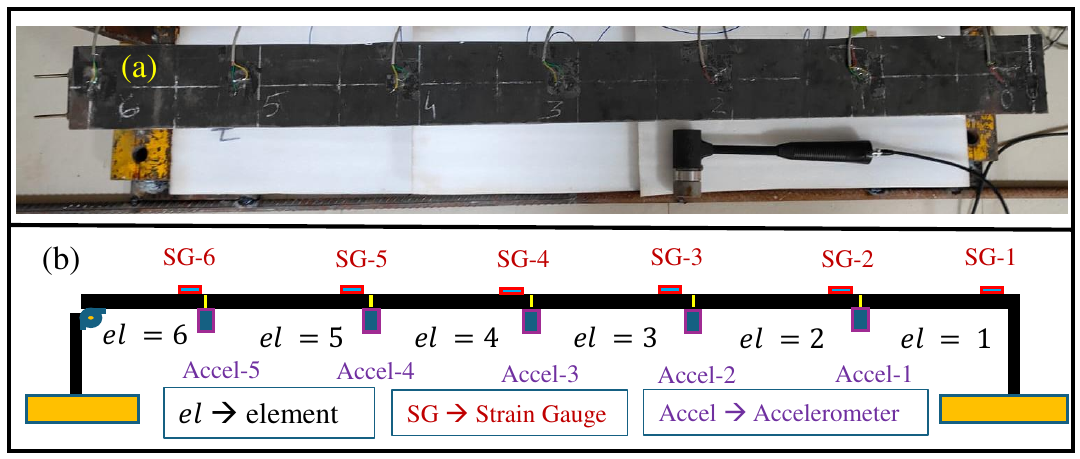}
	   \caption{Laboratory-scale experiment on a beam: (a) beam with impact hammer, (b) schematic diagram of the beam along with sensor positions }
	\label{FIG:beam_exp}
\end{figure}

\begin{figure}[b!]
	\centering
		\includegraphics[scale=.65]{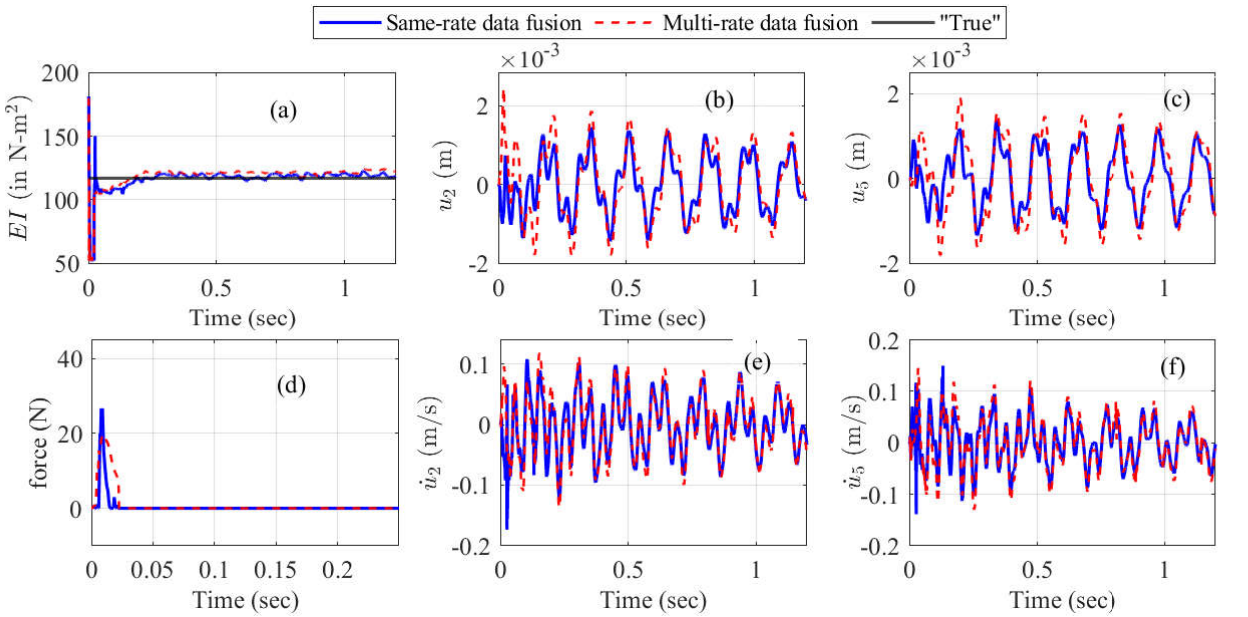}
	   \caption{(a) Comparison of estimated (a) parameter ($EI$) (d) input force, and (b-c, e-f) states (displacement and velocity) of the beam using the fused data of same-rate and multi-rate data}
	\label{FIG:beam_exp_ID}
\end{figure}

\subsection{Euler-Bernoulli Beam}
To validate the proposed methodology experimentally with strain-acceleration data fusion, an impact hammer test has been conducted on a laboratory-scale Euler-Bernoulli beam made of mild steel, as shown in Fig.~\ref{FIG:beam_exp}a, at the Structural Engineering Laboratory at IIT Kanpur, India. The instrumentation is shown in the schematic diagram of the beam in Fig.~\ref{FIG:beam_exp}b. The modulus of elasticity of the material is obtained by coupon test and found to be 180 MPa. The flexural rigidity ($EI$) of the beam, calculated using the coupon test results and the measured dimensions, is considered as the "true" value, and is equal to 117 N-m$^2$. 
\par
The impact hammer vibration test data was recorded with a sampling frequency of 2000 Hz for both acceleration and bending strain.
For validation purposes, two cases are considered: (i) the strain data downsampled to 200 Hz for multi-rate data fusion, and (ii) the recorded strain data at 2000 Hz for same-rate data fusion, with the acceleration at 2000 Hz in both cases. Incomplete instrumentation is considered in both cases with data from accelerometers: 1-2-3 and strain gauges: 1-4-6.
As the input is unknown in this case, both the input and the parameters are incorporated as unknowns within the state vector, referred to as the augmented state-parameter-input vector. The fused data is used to estimate the states and parameters along with the input, using the proposed approach. The augmented vector is considered as, $\mathbf{X} = \left\{u_1,u^\theta_1,u_2,\cdots,u^\theta_6,\dot{u}_1,\dot{u}^\theta_1,\dot{u}_2,\cdots,\dot{u}^\theta_6,f^t_2,EI,\alpha,\beta \right\}^\textrm{T}$, where $u^\theta_i$ and $\dot{u}^\theta_i$ are the rotational displacements and rotational velocities of the $i$th rotational DOF, and $f^t_2$ is the input force at the position of the accelerometer-2. Consequently, the state-space equation and the measurement function, as shown in the flowchart in Fig.~\ref{FIG:UKF_algo}, can be expressed solely in terms of this augmented vector.
In this case, the state-space equation becomes: $\mathbf{X}_{k+1} = \mathcal{F}(\mathbf{X}_{k})$.
Typically, a dual filtering approach is often employed to estimate states, parameters, and inputs simultaneously~\cite{azam2015dual,dertimanis2019input}. However, in this particular example, a single UKF-based algorithm sufficed to accurately estimate the parameters.
Herein, the estimation algorithm used is the CGUKF. From Fig.~\ref{FIG:beam_exp_ID}, it is evident that the estimation of $EI$, input force, and the states (i.e., displacement and velocity) using multi-rate fused data matches quite well with the estimation of those utilizing the same-rate fused data.

\section{Conclusion}
In structural monitoring, sampling frequencies of the recorded data can vary between different sensors (measuring the same or different physical quantities). This paper presents a direct methodology using UKF-based algorithms to estimate the states (i.e., displacements and velocities) as well as the structural parameters while utilizing multi-rate fused data. The proposed method involves a simple modification to the measurement update in UKF. The proposed modification can be used with any UKF-based algorithms.
The advantages of this methodology include: (a) Data fusion of high sampling frequency acceleration measurements can be performed with low sampling frequency measurements of both displacement and/or strain in the same framework; (b)~No~explicit data fusion algorithms are required in the pre-processing step; (c) The non-requirement of collocated sensing; (d) The states can be estimated at locations where sensors are not placed, along with the structural parameters; and (e) Simplicity and ease of implementation. 

The methodology is demonstrated through numerical illustrations with two different types of structures: a two-story shear building and a 2D truss. For the former structure, the acceleration responses are fused with displacement responses, while in the latter one, the acceleration responses are fused with strain responses.~In~both~cases, the accelerations are sampled at a higher frequency than the other response types. The methodology is validated with two laboratory-scale experimental applications: a four-storied steel frame tested on a shaking table and an impact hammer test on a Euler-Bernoulli steel beam. Accelerations, sampled at a higher frequency for both experimental datasets, are fused with displacements for the shake table test and strains for the impact hammer test. It is shown that the states and parameters are estimated with good accuracy, even without collocated sensing, and that the accuracy of the estimated states are comparable with existing data fusion techniques (which need collocated sensing).

\section{Data Availability Statement}
The data that support the findings of this study are available from the corresponding author upon reasonable request.
\section{Acknowledgments}
Financial support for this work has been provided by the Science and Engineering Research Board (SERB, DST, India), under grant numbers ECR/2017/003430 \& SERB/CE/2023789. The financial support is gratefully acknowledged. The authors sincerely thank Prof. Manolis Chatzis (Oxford University) for allowing use of the shake-table experimental data, and Prof. Samit Ray Chaudhuri (IIT Kanpur) for the help provided to conduct the experiment on the Euler-Bernoulli beam.
 
\printcredits

\bibliographystyle{model1-num-names}

\bibliography{references}


\end{document}